\documentclass[preprint,12pt]{elsarticle}
\PassOptionsToPackage{cmyk}{xcolor}
\usepackage[tbtags]{amsmath}
\usepackage{amssymb,bm,dcolumn,booktabs,threeparttable,multirow}
\usepackage[dvipsnames]{xcolor}
\definecolor{LinkBlue}{RGB}{6,69,173}
\definecolor{DarkBlue}{RGB}{11,0,128}
\usepackage[colorlinks=true,linkcolor=LinkBlue,urlcolor=LinkBlue,
	citecolor=LinkBlue,citecolor=LinkBlue]{hyperref}
\usepackage{graphicx,paralist}
\usepackage{aas_macros}
\setdefaultitem{}{\textbullet}{$\star$}{}

\journal{The Journal of High Energy Astrophysics}
\begin{document}
\begin{frontmatter}

\title{Cosmological evolution of primordial black holes}

\author{Jared R. Rice}
\ead{jrice@physics.unlv.edu}

\author{Bing Zhang}
\ead{zhang@physics.unlv.edu}

\address{Department of Physics and Astronomy, University of Nevada, Las Vegas, NV 89154, USA}

\begin{abstract}
The cosmological evolution of primordial black holes (PBHs) is considered. A comprehensive view of the accretion and evaporation histories of PBHs across the entire cosmic history is presented, with focus on the critical mass holes. The critical mass of a PBH for current era evaporation is $M_{cr}\sim 5.1\times10^{14}$ g. Across cosmic time such a black hole will not accrete radiation or matter in sufficient quantity to hasten the inevitable evaporation, if the black hole remains within an average volume of the universe. The accretion rate onto PBHs is most sensitive to the mass of the hole, the sound speed in the cosmological fluid, and the energy density of the accreted components. It is not easy for a PBH to accrete the average cosmological fluid to reach $30M_\odot$ by $z\sim0.1$, the approximate mass and redshift of the merging BHs that were the sources of the gravitational wave events GW150914 and GW151226. A PBH located in an overdense region can undergo enhanced accretion leading to the possibility of growing by many orders of magnitude across cosmic history. Thus, two merging PBHs are a plausible source for the observed gravitational wave events. However, it is difficult for isolated PBHs to grow to supermassive black holes (SMBHs) at high redshift with masses large enough to fit observational constraints.
\end{abstract}


\end{frontmatter}

\section{\label{sec:Intro} Introduction}
\noindent Primordial black holes (PBHs) are among the most intriguing ghosts in the universe. A singular PBH of sufficient mass can navigate the history of the universe without detectable clues to its existence; a true cosmic ghost. Low mass PBHs evaporate before the current epoch and the radiation signature of an isolated high mass PBH is too weak to detect. The last moments of a PBH evaporation reveal the hole through a burst of high-energy radiation that is distinguishable from that of short gamma-ray bursts (GRBs) \cite{Ukwa16}.\\
\indent The upper limits on the number density of PBHs across a wide range of masses is discussed extensively in \cite{Carr10,Carr16}. To date there are no confirmed PBH burst signals, but these compelling ghosts are ripe cosmological messengers that will enhance our understanding of the universe if observed. The PBHs evolving through cosmic history could be used as a proxy for understanding the conditions in the early universe. PBHs of significant mass may gain a dark matter (DM) halo, e.g. \cite{Mack07,Rico08}. Since the PBH evaporation rate depends only on the mass of the hole and the assumed particle physics model \cite{Ukwa15}, PBHs in similar astrophysical environments should produce similar radiation signatures; the ultimate ``standard candles.''\\
\indent This study explores the evolution of PBHs through accretion and evaporation across the entire cosmic history. Special attention is paid to the changes in the density, temperature, and sound speed in the cosmological fluid because of their influence on the accretion rate of that fluid onto the PBHs. In \S\ref{sec:Cosmo} the concordance cosmological model of $\Lambda$CDM is discussed. In \S\ref{sec:Evapo} the PBH accretion and evaporation models are discussed and formulae are given for the accretion rates in the various cosmological eras. In \S\ref{sec:Results} the results of the study are discussed. Finally in \S\ref{sec:Discussion} the conclusions are presented and a discussion of astrophysical implications is made.

\section{\label{sec:Cosmo} Cosmological model}
\noindent The concordance cosmology assumed throughout this study is the six parameter $\Lambda$CDM model, implementing the most recent \emph{Planck} Collaboration results \cite{Plan15}. The model consists of the homogeneous and isotropic Friedmann-Robertson-Walker (FRW) geometry dynamically evolving according to the Einstein field equations. The Einstein equations, also called the Friedmann equations \cite{Frie22,Hawk73} in this case, describe the evolution of the curvature and energy content of the universe as
\begin{align}
\biggl(&\frac{\dot{a}}{a}\biggr)^2 +\frac{kc^2}{a^2} = \frac{8\pi G}{3}\rho,\label{eq:Fried_1}\\
&\frac{\ddot{a}}{a} = -\frac{4\pi G}{3}\biggl(\rho+\frac{3P}{c^2}\biggr),\label{eq:Fried_2}
\end{align}
\noindent where the scale factor is $a\equiv a_0(1+z)^{-1}$, with the scale factor today $a_0\equiv1$ and $z$ the cosmological redshift, $k=0,\pm1$ indicating zero, positive, or negative spatial curvature respectively, $G$ is the universal gravitation constant, and $c$ is the speed of light. The term $\rho$ is the sum of the proper inertial mass densities of the cosmological fluid and the contribution from spatial curvature, and $P$ is the pressure contribution from matter, radiation, and vacuum energy (or cosmological constant $\Lambda$).\\
\indent The equation of state of each cosmological fluid can be expressed $P_i = w_i\rho_i c^2$ (no sum over $i$) with equation of state parameter $w_i$. The equation of state parameters for matter, radiation, cosmological constant, and spatial curvature are $0$, $1/3$, $-1$, and $-1/3$ respectively. Note that the equation of state of a baryonic gas $P\propto\rho^\gamma$, where $\gamma$ is the adiabatic index, is relevant when calculating accretion rates onto a compact object. The approximation $w_{i=b}\approx0$ for a baryonic gas holds on cosmological scales.\\
\indent The Hubble parameter $H$ is a measure of the temporal (extrinsic) curvature of the FRW geometry and is defined
\begin{align}
\biggl(\frac{\dot{a}}{a}\biggr)^2 &\equiv H^2 = H_0^2\mathcal{E}^2,\label{eq:Fried}
\end{align}
\noindent where subscript-0 implies evaluation of a quantity today. The Hubble constant is $H_0 = 100h\text{ km\,s$^{-1}$Mpc$^{-1}$}$ where the dimensionless Hubble parameter is $h=0.6774$ from \emph{Planck} \cite{Plan15}. Before defining $\mathcal{E}^2$, it is convenient to introduce the dimensionless density parameters today
\begin{align} \Omega_{i,0} &\equiv \frac{\rho_{i,0}}{\rho_{cr,0}} = \frac{8\pi G}{3H_0^2}\rho_{i,0},\label{eq:parameters}
\end{align}
\noindent where $i$ indicates baryonic matter, dark matter, radiation, and $\Lambda$. Dividing Eq. (\ref{eq:Fried_1}) by $H^2$ and evaluating the quantities today gives an expression for the `effective' dimensionless density parameter for spatial curvature
\begin{align}
\Omega_{k,0} &= 1 - \Omega_0,\label{eq:Omega_k}
\end{align}
\noindent where $\Omega_0 \equiv \Omega_{r,0}+\Omega_{m,0}+\Omega_{\Lambda,0}$. The combined \emph{Planck} and baryon acoustic oscillation data \cite{Plan15} are consistent with $\Omega_{k,0} = 0.000\pm0.005$, i.e. the universe has zero spatial curvature to within $0.5\%$ accuracy. The term $\mathcal{E}^2$ in Eq. (\ref{eq:Fried}) is a function of the dimensionless density parameters with their redshift dependencies
\begin{align}
\nonumber\mathcal{E}^2 &\equiv \sum_i\Omega_{i,0}(1+z)^{3(1+w_i)}\\
\begin{split}
&=\Omega_{r,0}(1+z)^4 + \Omega_{m,0}(1+z)^3\\
&\phantom{==}+(1-\Omega_0)(1+z)^2 + \Omega_{\Lambda,0}.\label{eq:Hubble}
\end{split}
\end{align}
\indent In the FRW geometry, proper time is related to redshift through the differential $\dot{z} = -H(z)(1+z)$. Therefore the time $\Delta t\equiv t_2-t_1$ elapsed between any two redshifts $z_1$ and $z_2$ is given by the integral
\begin{align}
\Delta t &\equiv \int_{t_1}^{t_2} dt = H_0^{-1}\int_{z_2}^{z_1}\frac{dz}{(1+z)\mathcal{E}(z)}\label{eq:time},
\end{align}
\noindent which has no tractable analytic solution ordinarily, but may be calculated analytically in simple cases or numerically in general. The age of the universe calculated numerically from Eq. (\ref{eq:time}) is $t_0=13.8$ Gyr, which was reported in the 2015 \emph{Planck} results \cite{Plan15}.\\
\indent The spatially-averaged inertial mass densities of the various components of the cosmological fluid decrease as power-laws with decreasing redshift according to their equation of state, i.e.
\begin{align}
\rho_i &\propto (1+z)^{3(1+w_i)}\label{eq:rho_evol}
\end{align}
\noindent for matter, radiation, curvature, and the cosmological constant. The average matter density in the universe evolves as $\rho_m \propto (1+z)^3$. The effective mass density of radiation evolves as $\rho_r \propto (1+z)^4$. In the early universe the redshift dependence of this term is more complicated due to the presence of radiation in the form of neutrinos and other relativistic Standard Model (SM) particles in addition to photons. Thus $\rho_r$  evolves \cite{Kolb90,Jung96} according to the expression
\begin{align}
\rho_r(T) &= \frac{\pi^2}{30}g_\star(T)\frac{k_B^4T^4}{c^5\hbar^3},\label{eq:rho_rad}
\end{align}
\noindent where $g_\star(T)$ is the effective number of relativistic degrees of freedom, $k_B$ is Boltzmann's constant, and $T$ is the temperature. Thus the effective radiation mass density evolves with redshift as $\rho_r\propto g_\star(1+z)^4$.
\begin{table}
\begin{center}
\caption{\label{tab:particles} Standard model elementary particles and other mass thresholds important in the early universe. Listed are the particles that freeze out below each temperature threshold, the mass of each particle or threshold from \cite{Oliv14}, the effective relativistic degrees of freedom at the temperature corresponding to that mass threshold, and the change in the degrees of freedom as the radiation temperature of the universe crosses the threshold. See also Fig. \ref{fig:gstar}.}
\begin{threeparttable}
\renewcommand{\arraystretch}{1.2}		
\begin{tabular}{cccr}
\toprule
Particle(s) & Mass [MeV]& $g_\star$\tnote{a}& $-\Delta g_\star$\\
\hline
All & $>m_{t,\bar{t}}$ & $\!\!\!106.75$ & ---\\
$t,\bar{t}$ & $1.73\times10^5$ & $96.25$ & $\frac{7}{8}\cdot2\cdot2\cdot3$\\
$H^0$ &$1.26\times10^5$ & $95.25$ & $1$\\
$Z^0$ &$9.12\times10^4$ & $92.25$ & $3$\\
$W^\pm$ &$8.04\times10^4$ & $86.25$ & $2\cdot3$\\
$b,\bar{b}$ &$4.18\times10^3$ & $75.75$ & $\frac{7}{8}\cdot2\cdot2\cdot3$\\
$\tau^\pm$ &$1.78\times10^3$ & $72.25$ & $\frac{7}{8}\cdot2\cdot2$\\
$c,\bar{c}$ &$1.28\times10^3$ & $61.75$ & $\frac{7}{8}\cdot2\cdot2\cdot3$\\
$\Lambda_\text{QCD}\tnote{b}$ & $170$ & $17.25$ & $44.5$\\
$\pi^\pm$ &$140$ & $15.25$ & $2\cdot1$\\
$\pi^0$ &$135$ & $14.25$ & $1\cdot1$\\
$\mu^\pm$ &$106$ & $10.75$ & $\frac{7}{8}\cdot2\cdot2$\\
$\nu_{dec}$\tnote{c} &$2.6$ & $\phantom{1}7.25$ & $\frac{7}{8}\cdot2\cdot2$\\
$e^\pm$ &$0.511$ & $\;\:\,\,3.38$\tnote{d} & $-\Delta g_{\star,f}$\tnote{e}\\
\bottomrule
\end{tabular}
\begin{tablenotes}
\item[a]{$g_\star$ at or below corresponding mass threshold}
\item[b]{QCD phase transition \cite{Gupt11}; remaining quarks ($s\bar{s}$, $d\bar{d}$, $u\bar{u}$) and gluons are bound in hadrons}
\item[c]{Neutrino decoupling energy threshold}
\item[d]{$g_\star(T<m_e) = 2+\frac{7}{8}\cdot2\cdot N_\text{eff}\cdot(4/11)^{4/3}\sim3.38$; where $N_\text{eff}\sim 3.04$ \cite{Plan15}}
\item[e]{$-\Delta g_{\star,f} = \frac{7}{8}\cdot2\cdot3-\frac{7}{8}\cdot2\cdot N_\text{eff}\cdot(4/11)^{4/3}$}
\end{tablenotes}
\end{threeparttable}
\end{center}
\end{table}

\begin{figure}[p]
\begin{center}
\includegraphics[width=\columnwidth]{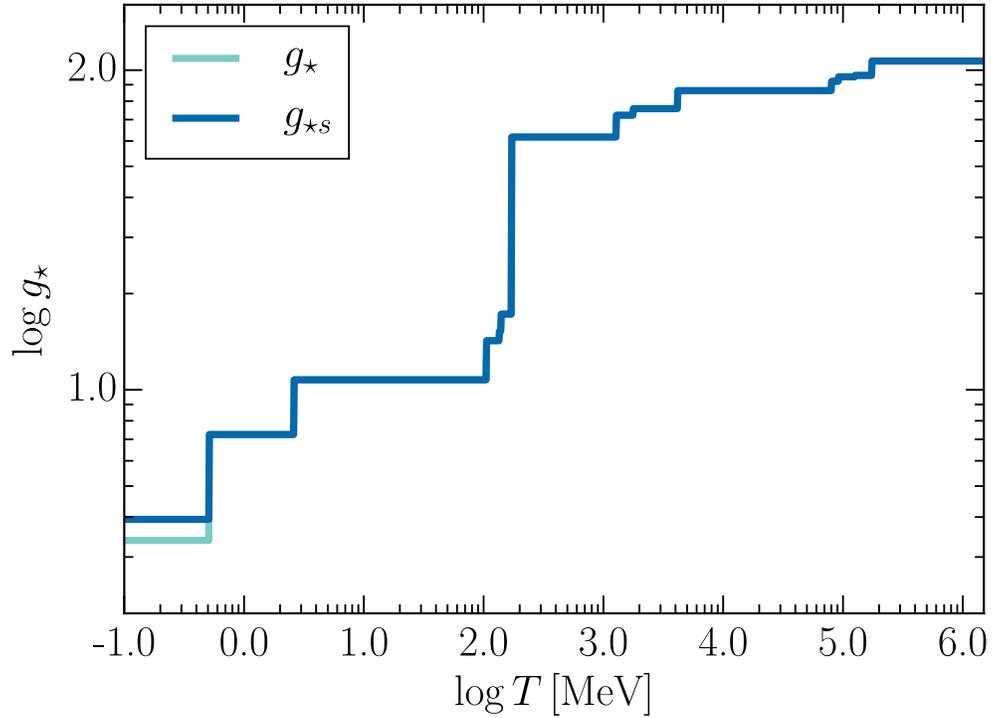}
\caption{\label{fig:gstar} Stepwise approximation of $g_\star$ as a function of temperature, including only relativistic particles whose number density is high enough to contribute. Steps occur at temperatures corresponding to the rest mass of elementary SM particles with the largest step occuring at the QCD phase transition scale $\Lambda_\text{QCD}\sim 170$ MeV \cite{Gupt11}. The value of $g_\star$ is a minimum for temperatures less than the neutrino decoupling temperature and a maximum for temperatures greater than or equal to the top quark mass. See \cite{Kolb90} for a discussion of $g_{\star s}$.}
\end{center}
\end{figure}

\indent A list of the important particle mass and energy thresholds is given in Table \ref{tab:particles}. Shown in Fig. \ref{fig:gstar} is the corresponding plot of $g_\star$ as a function of temperature.
\noindent The factor $g_\star$ increases by up to a factor $106.75/3.38\sim31.6$ at high redshift when all SM particles are relativistic. The factor $g_\star(T)$ for all relativistic particle species in thermal equilibrium can be calculated as the sum \cite{Kolb90}
\begin{align}
g_\star(T) &= \sum_{i=b}g_i\biggl(\frac{T_i}{T}\biggr)^4 + \frac{7}{8}\sum_{i=f}g_i\biggl(\frac{T_i}{T}\biggr)^4,
\end{align}
\noindent where the first term is the sum over all bosons and the second term is the sum over all fermions.\\
\indent Each component of the cosmological fluid has an associated temperature whose value depends on redshift. At high redshift after inflation the universe is dominated by radiation and the components of the cosmological fluid are in equilibrium. The temperature of radiation evolves simply as $T_r = T_0(1+z)$ where $T_r$ is the radiation temperature at any redshift lower than the neutrino decoupling redshift $z_{dec,\nu}$ and the temperature today is $T_0 = 2.72548\text{ K}$ \cite{Fixs09}. The baryonic matter is coupled to the radiation through Compton scattering prior to the thermalization redshift \cite{Peeb93}
\begin{align}
1 + z_{th} &\sim 800\bigl(\Omega_{b,0}h^2\bigr)^{2/5} \sim 174,
\end{align}
\noindent and in this redshift regime the baryonic matter temperature evolves as $T_b = T_0(1+z)$. After $z_{th}$, the baryonic matter temperature evolves adiabatically until reionization as
\begin{align}
T_b = T_0\frac{(1+z)^2}{1+z_{th}},
\end{align}
\noindent so that at thermalization, $T_b = T_r(z=z_{th})$.\\
\indent It is assumed that the dark matter is mostly weakly interacting massive particles (WIMPs) denoted $\chi$. For simplicity other dark matter models are not considered in this study; see \cite{Feng10} for more details on all the dark matter models. For a discussion on the possibility of the PBHs themselves being the dark matter see \cite{Carr10}. If the dark matter is composed of WIMPs the DM temperature decouples from the radiation temperature at high redshift when thermal freeze out of the dark matter particles occurs \cite{Feng10}. The model forces the DM thermal freeze out to occur when $k_BT_{fr}\sim m_\chi c^2/20$. If $m_\chi c^2 = 100\text{ GeV}$ the freeze out redshift is $1+z_{fr} = T_{fr}/T_0\sim2\times10^{13}$.\\
\indent The root-mean-square velocity $v_{rms}$ is approximated by comparing the relativistic kinetic energy of the dark matter particles to their thermal energy:
\begin{align}
(\gamma_\chi-1)m_\chi c^2 &= \frac{3}{2}k_BT_\chi, \label{eq:dmat_rms}
\end{align}
\noindent where the Lorentz factor associated with $\beta_\chi \equiv v_{rms}/c$ of the dark matter particles is given by
\begin{align}
\gamma_\chi &\equiv (1-\beta_\chi^2)^{-1/2}.\label{eq:DM_Lorentz}
\end{align}
\noindent It is useful to define the dimensionless quantity
\begin{align}
\Theta_\chi &\equiv\frac{3k_BT_\chi}{2m_\chi c^2}
\end{align}
\noindent such that Eq. (\ref{eq:dmat_rms}) becomes $\gamma_\chi = 1+\Theta_\chi$ and therefore
\begin{align}
\beta_\chi^2 &= \frac{\Theta_\chi^2+2\Theta_\chi}{(1+\Theta_\chi)^2}, \label{eq:v_rms}
\end{align}
\noindent which is friendly to numerical evaluations for all possible physical values of $\Theta_\chi$. The value of $\beta_\chi$ becomes close to unity quite rapidly and for the assumed freeze-out temperature, $\beta_\chi \simeq 0.367$, i.e. the dark matter particles are mildly relativistic at freeze-out.

\section{\label{sec:Evapo} Primordial black hole accretion and evaporation}
\noindent The history of the universe may be divided into redshift regimes to simplify analysis. The relevant physical processes in the very early universe are distinct from those acting in the current era and thus it is important to summarize the physics in each regime.\\
\indent Though important to the dynamics of the universe in general, the history of the very early universe (prior to inflation) is not considered in detail in this study. The number density of a pre-inflation cosmological relic, e.g. any pre-inflation PBHs, is negligible after the inflationary epoch. The number density of a pre-inflation relic depends on the amount of inflation. The amount of inflation is calculated by finding the number of $e$-foldings during the inflationary epoch
\begin{align}
N &\equiv \int_{a_i}^{a_f}d\ln a = \int_{t_i}^{t_f} H\,dt,
\end{align}
\noindent where $t_i$ denotes the start of inflation, $t_f$ denotes the end of inflation, and $H$ is the Hubble parameter given in Eq. (\ref{eq:Fried}). A successful inflation model requires the number of $e$-foldings to be at least $N_{min}\sim50$ in order to solve the horizon problem \cite{Remm14}. The number density of a relic which formed prior to inflation will thus decrease by a factor $e^{3N_{min}}\sim e^{150} \sim 10^{65}$. Thus any PBHs which formed prior to inflation are unlikely to be located in the observable universe today.\\
\indent After inflation, PBHs may form through a variety of mechanisms including collapse of primordial inhomogeneities, phase transitions, and cosmic string or domain wall collisions \cite{Carr10}. If the energy density fluctuations have a strength $\delta\rho/\rho\sim 1$ in a particular spacetime volume, the region will likely collapse to a black hole. In this study it is assumed that the collapse to a black hole occurs on a time scale much shorter than the Hubble time so that the expansion is irrelevant to PBH formation. A black hole forming at a time $t$ after the Big Bang will have a mass less than or equal to the Hubble mass at that time, i.e.
\begin{align}
M_{H} &= \frac{c^3t}{G}\sim (4.0\times10^{14}\text{ g})t_{-24},\label{eq:Hubble_mass}
\end{align}
\noindent using the useful notation $f = 10^nf_n$. At $t=1.0$ s after the Big Bang, the Hubble mass is $M_H \sim 2.0\times10^5M_\odot$. A derivation of Eq. (\ref{eq:Hubble_mass}) can be found in \ref{Appendix_A}. Models for the mass function of PBHs are discussed in \cite{Carr10} with emphasis on the behavior of the mass function near the critical mass regime.\\
\indent Particles with spin $s$ between energy $E$ and $E+dE$ are emitted near the horizon of a Schwarzschild black hole of mass $M$ at a rate \cite{Hawk75,MacG91}
\begin{align}
d\dot{N} &= \frac{\Gamma_sdE}{2\pi\hbar}\biggl[\exp\biggl(\frac{8\pi GME}{\hbar c^3}\biggr)-(-1)^{2s}\biggr]^{-1},\label{eq:HawkRate}
\end{align}
\noindent where $\Gamma_s$ is the absorption probability for a mode with spin $s$ \cite{Page76}. This is the so-called Hawking radiation which resembles emission from a blackbody with radius $R_s = 2GM/c^2$ of temperature
\begin{align}
k_BT_{BH} &= \frac{\hbar c^3}{8\pi GM} = (10.6\text{ MeV})M_{15}^{-1}.\label{eq:HawkTemp}
\end{align}
\noindent The BH can only radiate when the temperature of the hole is greater than that of the radiation bath of the early universe. The temperature of radiation in the early universe evolves as $T_r\propto 1+z$ and is less than $10.6$ MeV when $z<4.5\times10^{10}$, i.e. when $t>0.01$ s. This will have a negligible effect on the evolution of PBHs near $10^{15}$ g. The absorption probability asymptotes to the geometric optics limit
\begin{align}
\Gamma_s &= \frac{27G^2M^2E^2}{\hbar^2c^6}\label{eq:geometric_optics}
\end{align}
\noindent when the particle energy is  $E \gg k_BT_{BH}$. The functional form of $\Gamma_s$ is much more complicated for lower energy $E\sim k_BT_{BH}$ interactions as discussed in \cite{Page76}.\\
\indent The mass loss rate due to the Hawking emission from a Schwarzschild black hole of mass $M$ requires a sum over all particle types and an integration over the particle energies leading to the simple equation 
\begin{align}
\frac{dM_{15}}{dt} &= (-5.34\times10^{-5}\text{ g$\,$s$^{-1}$})f(M)M_{15}^{-2},\label{eq:Hawk}
\end{align}
\noindent where $f(M)$ is a function \cite{MacG91} allowing for the emission of particles other than photons and $f(M)=1$ for $M\gg10^{17}$ g. The function $f(M)$ increases when the mass of the PBH crosses a particle mass threshold (see Table \ref{tab:particles}), after which the PBH may emit that particle. A good approximation is $f(M) \sim f(M_i)$ because for the majority of its lifetime the mass of a PBH remains near its formation mass $M_i$ \cite{Page76}.\\
\indent For supermassive black holes or stellar mass black holes the evaporation rate in Eq. (\ref{eq:Hawk}) is negligibly small. The evaporation rate becomes important on cosmological time scales for black holes with mass $M\sim10^{15}$ g. This is seen by integrating Eq. (\ref{eq:Hawk}) to get the evaporation timescale
\begin{align}
t_{evap} &= (6.24\times 10^{18}\text{ s})f(M_i)^{-1}M_{i,15}^3.\label{eq:evaptime}
\end{align}
\noindent Assuming $t_{evap} = t_0 = 13.8$ Gyr, the critical formation mass for evaporation today is
\begin{align}
M_{cr} &= 5.1\times 10^{14}\text{ g},\label{eq:mass_crit}
\end{align}
\noindent where the parameter $f$ for the critical mass is $f(M_{cr})\sim1.9$ as assumed in \cite{Carr10,MacG91}.\\
\indent In every cosmic era the Hawking evaporation of a near-critical mass PBH will compete with accretion of the cosmological fluid onto the hole. For the PBHs of $M\sim M_{cr}$ the accretion turns out to be irrelevant if the hole accretes the cosmological fluid at spatially-averaged densities. For PBHs much smaller than $M_{cr}$ accretion is completely unimportant. For PBHs larger than $M_{cr}$ the accretion becomes ever more important and the evaporation rate becomes ever smaller. Thus it is important to quantify the various accretion rates at the relevant cosmic epochs. The accretion rates are dependent on the physical parameters of the cosmological fluid, which change dramatically with redshift. The full equation to be solved is the first-order nonlinear ordinary differential equation in $M$
\begin{align}
\frac{dM}{dt} &= \dot{M}_{evap}(f,M;z) + \dot{M}_{acc}(\rho,c_s,M;z),\label{eq:diff_eq}
\end{align}
\noindent where $\dot{M}_{evap}$ is given by Eq. (\ref{eq:Hawk}) and the mass accretion term $\dot{M}_{acc}$ will be calculated in the following sections. The equation is integrated from the formation time $t_i$ to any desired final time (or the evaporation time for small holes) using Eq. (\ref{eq:time}), with the concordance cosmological model accounted for at all times. The Hawking evaporation term has an explicit dependence on $f(M)$ and the mass $M$ of the hole and an implicit dependence on $z$ (or $t$). The accretion term has explicit dependence on $\rho$, $c_s$, and $M$ and an implicit dependence on $z$ due to the evolution of those quantities across cosmic time.\\
\indent The mass accretion term in Eq. (\ref{eq:diff_eq}) is split into its component parts
\begin{align}
\dot{M}_{acc}(\rho,c_s,M;z) &= \dot{M}_{r} + \dot{M}_{b} + \dot{M}_\chi,\label{eq:mass_acc}
\end{align}
\noindent where $r$ indicates radiation ($\gamma$, $\nu$, and other SM particles), $b$ indicates baryonic matter, and $\chi$ the dark matter particles. When the universe is cool enough (i.e. $T_r < 0.511$ MeV), the radiation term consists only of photons and neutrinos. At higher redshift, the other SM particles become relativistic and can be accreted. When the baryonic matter is coupled to the radiation the two accretion rates become coupled and are written $\dot{M}_{b+r}$. In the sections following, the mass accretion term is calculated explicitly for the different cosmic eras.

\subsection{\label{subsec:Late} Late universe accretion}
\noindent In the late universe at $z \lesssim z_{th}$ the relevant cosmic scales are set by the formation and evolution of structure, i.e. the distribution of dark and baryonic matter in the cosmic web. The details of cosmic structure formation are ripe with rich and complicated physics and are not included in this study; see \cite{Prim15,Mo2010}.\\
\indent To set a bound on late universe accretion, all accretion terms here are set by the spatially-averaged fluid quantities. The PBHs in our universe will likely form and evolve within overdense regions, so the use of spatially-averaged quantities gives a good idea what to expect with relatively isolated holes. The accretion of radiation in this redshift regime is unimportant for $\sim M_{cr}$ PBHs because it is a horizon-limited growth given in the approximate form
\begin{align}
\nonumber\dot{M}_r &= 4\pi R_S^2c\rho_r\\
\nonumber&= \frac{16\pi G^2}{c^3}\rho_{r,0}(1+z)^4M^2\\
&= (6.5\times10^{-48}\text{ g$\,$s$^{-1}$})(1+z)^4M_{15}^2,\label{eq:rad_late}
\end{align}
\noindent where $\rho_r$ is the equivalent mass density in radiation. The accretion rate in Eq. (\ref{eq:rad_late}) is comparable to the magnitude of the Hawking evaporation rate when the mass of the PBH is
\begin{align}
M &= (6.3\times10^{25}\text{ g})(1+z)^{-1}.\label{eq:rad_late_crit}
\end{align}
\noindent Thus the accretion of background radiation in this redshift regime is unimportant to critical mass holes. A PBH of the mass given in Eq. (\ref{eq:rad_late_crit}) will not evaporate until long after the current era.\\
\indent The accretion of baryonic matter is more complicated as it is governed by gas dynamics in the vicinity of the PBH. If the PBH is in an `average' region of the universe, i.e. of average baryonic matter density and temperature, the accretion of baryons will be a competition between Bondi and Eddington-limited accretion \cite{Bond52}. The accretion rate found in this manner will inform a lower bound for any relevant PBH accretion activity. Below $z\sim 30$ the details of cosmic structure will change this simplified picture, but it is useful to set a first approximation. A complete picture of the baryonic accretion has not been properly solved and is the subject of intense study from both theoretical and observational perspectives. With the complicated gas dynamics removed from the analysis in this simplified calculation, the accretion rate can be expressed as
\begin{align}
\nonumber\dot{M}_b &= \min\bigl(\dot{M}_{b,B},\dot{M}_{b,E}\bigr)\\
&= \min\biggl(\frac{4\pi\lambda_sG^2}{c_{s,b}^3}\rho_{b,0}(1+z)^3M^2,\frac{4\pi Gm_p}{\sigma_Tc}M\biggr).\label{eq:late_Bond_Edd}
\end{align}
\noindent where $\lambda_s = 1/4$ for a $\gamma = 5/3$ baryonic gas \cite{Bond52}, $m_p$ is the proton mass, and $\sigma_T$ is the Thomson scattering cross section for electrons. It is clear that the Eddington limit is redshift-independent and is equal to
\begin{align}
\dot{M}_E &= (7.03\times10^{-2}\text{ g$\,$s$^{-1}$})M_{15}.\label{eq:Edd}
\end{align}
\noindent The Bondi rate is also redshift-independent in this redshift regime. The temperature of the baryonic gas in this regime is
\begin{align}
T_b &= \frac{T_r(1+z)}{1+z_{th}} = \frac{T_0(1+z)^2}{1+z_{th}},\label{eq:baryon_temp}
\end{align}
\noindent and therefore the sound speed in the baryonic gas (assuming it is entirely hydrogen) is
\begin{align}
\nonumber c_{s,b} &= \biggl(\frac{5k_BT_b}{3m_p}\biggr)^{1/2}\\
\nonumber &= (1+z)\biggl(\frac{5k_BT_0}{3m_p(1+z_{th})}\biggr)^{1/2}\\
&= (1.5\times10^3\text{ cm$\,$s$^{-1}$})(1+z).\label{eq:baryon_sound}
\end{align}
\noindent Thus the redshift dependence of the Bondi rate goes away and Eq. (\ref{eq:late_Bond_Edd}) becomes
\begin{align}
\nonumber\dot{M}_b &= \min\biggl[\pi G^2\biggl(\frac{5k_BT_0}{3m_p(1+z_{th})}\biggr)^{-3/2}\rho_{b,0}M^2,\frac{4\pi Gm_p}{\sigma_Tc}M\biggr]\\
&= \begin{cases}
(1.9\times10^{-24}\text{ g$\,$s$^{-1}$})M_{15}^2, & M<M_{cr,1}\\
(7.0\times10^{-2}\text{ g$\,$s$^{-1}$})M_{15}, & M>M_{cr,1}\label{eq:Bond_min}
\end{cases},
\end{align}
\noindent where $M_{cr,1} = 3.8\times10^{37}$ g is the mass of a PBH that gives an equivalence in the Bondi and Eddington rates in this redshift regime. Comparing Eq. (\ref{eq:Bond_min}) to Eq. (\ref{eq:Hawk}) it is clear that an isolated near-critical mass PBH cannot accrete sufficiently to beat the Hawking evaporation rate. The Bondi rate is comparable to the magnitude of the Hawking evaporation rate when the PBH has the characteristic mass
\begin{align}
M_{ch,1} &= 7.3\times10^{19}\text{ g}.\label{eq:M_ch1}
\end{align}
\indent Any relevant growth of a near-critical mass PBH in this redshift regime will have to come from enhanced accretion if the hole is located within a significant density perturbation such as an individual galaxy or galaxy cluster.

\subsection{\label{subsec:Post}Post-recombination accretion}
\noindent In the post-recombination universe ($z_{th} < z < z_{rec}$) the matter temperature is coupled to the radiation temperature via Compton scattering, i.e. $T_b = T_0(1+z)$. The recombination redshift is listed in \cite{Plan15} as $z_{rec} = 1089.90$. Starting 381,000 yr after the Big Bang and until thermal decoupling, the sound speed in the baryonic gas can be expressed as
\begin{align}
\nonumber c_{s,b} &= \biggl(\frac{5k_BT_b}{3m_p}\biggr)^{1/2}\\
\nonumber &= (1+z)^{1/2}\biggl(\frac{5k_BT_0}{3m_p}\biggr)^{1/2}\\
&= (1.9\times10^4\text{ cm$\,$s$^{-1}$})(1+z)^{1/2}\label{eq:bary_sound_post}.
\end{align}
\indent In this redshift regime, the accretion of radiation is still horizon limited and given by Eq. (\ref{eq:rad_late}). The accretion of baryonic matter is the Bondi accretion rate at lower mass and is Eddington-limited growth if the mass is large enough. Using the same arguments as before
\begin{align}
\nonumber\dot{M}_b &= \min\bigl(\dot{M}_{b,B},\dot{M}_{b,E}\bigr)\\
\nonumber &= \min\biggl(\frac{\pi G^2}{c_{s,b}^3}\rho_{b,0}(1+z)^3M^2,\frac{4\pi Gm_p}{\sigma_Tc}M\biggr)\\
&= \begin{cases}
(8.1\times10^{-28}\text{ g$\,$s$^{-1}$})(1+z)^{3/2}M_{15}^2, & M<M_{cr,2}\\
(7.0\times10^{-2}\text{ g$\,$s$^{-1}$})M_{15}, & M>M_{cr,2}\label{eq:rec_Bond_Edd}
\end{cases},
\end{align}
\noindent where $M_{cr,2} = (8.7\times10^{40}\text{ g})(1+z)^{-3/2}$ is the PBH mass that gives an equivalent Bondi and Eddington rate. The Bondi accretion rate in Eq. (\ref{eq:rec_Bond_Edd}) is comparable to the magnitude of the Hawking evaporation rate when the PBH has a characteristic mass
\begin{align}
M_{ch,2} &= (5.1\times10^{20}\text{ g})(1+z)^{-3/8}.\label{eq:M_ch2}
\end{align}
\noindent Thus in the post-recombination era until thermal decoupling, the relevant process for near-critical mass PBHs is Hawking evaporation.

\subsection{\label{subsec:Pre} Pre-recombination accretion}
\noindent In the pre-recombination era ($z_{rec} < z < z_{mr}$) after matter-radiation equality the baryonic matter and radiation are fully coupled and cannot accrete independently. Thus the assumptions present in the Bondi accretion formula fail \cite{Carr81} and the accretion of the coupled fluid is horizon-limited. The temperature of the baryonic gas is coupled to the radiation temperature and the sound speed in the fluid can be written (see \ref{Appendix_B})
\begin{align}
c_s^2 &= \frac{c^2}{3}\frac{4\rho_r}{4\rho_r+3\rho_b}.\label{eq:sound_rad_bary}
\end{align}
\noindent At the matter-radiation equality the sound speed in Eq. (\ref{eq:sound_rad_bary}) is a few percent below the asymptotic value $c/\sqrt{3}$.\\
\indent The accretion rate of this coupled fluid onto a PBH is the horizon-limited rate
\begin{align}
\nonumber\dot{M}_{b+r} &= 4\pi R_S^2c_s(\rho_r+\rho_b)\\
\nonumber&= \frac{16\pi G^2}{\sqrt{3}c^3}\biggl[\frac{4\rho_{r,0}(1+z)^4}{4\rho_{r,0}(1+z)^4+3\rho_{b,0}(1+z)^3}\biggr]^{1/2}\\
&\phantom{==}\times\Bigl[\rho_{r,0}(1+z)^4+\rho_{b,0}(1+z)^3\Bigr]M^2.\label{eq:rad_bary}
\end{align}
\noindent This rate has a complicated dependence on redshift so it is useful to expand the right hand side of Eq. (\ref{eq:rad_bary}) near the boundaries of this redshift regime. Defining the intermediary terms $\rho_r'\equiv4\rho_{r,0}$ and $\rho_b'\equiv3\rho_{b,0}$ near $z_{rec}$ the rate takes the form
\begin{align}
\nonumber\dot{M}_{b+r} &= (6.8\times10^{-36}\text{ g$\,$s$^{-1}$})\Biggl[0.26\rho_r'\biggl(\frac{1+z}{1+z_{rec}}\biggr)^4\\
&\phantom{==}+0.17\rho_b'\biggl(\frac{1+z}{1+z_{rec}}\biggr)^3\Biggr]M_{15}^2.\label{eq:rad_bary_rec}
\end{align}
\noindent The redshift dependence of the sound speed in Eq. (\ref{eq:sound_rad_bary}) is included in the expansion above and in the expansion that follows. The rate in Eq. (\ref{eq:rad_bary_rec}) becomes comparable in magnitude to the Hawking evaporation rate when the PBH has a characteristic mass
\begin{align}
\nonumber M_{ch,3a} &\simeq (6.2\times10^{22}\text{ g})\Biggl[0.26\rho_r'\biggl(\frac{1+z}{1+z_{rec}}\biggr)^4\\
&\phantom{==}+0.17\rho_b'\biggl(\frac{1+z}{1+z_{rec}}\biggr)^3\Biggr]^{-1/4}.\label{eq:M_ch3a}
\end{align}
\noindent So again the Hawking evaporation is most important for critical mass, i.e. Eq. (\ref{eq:mass_crit}), PBHs. Closer to $z_{mr}$ the rate in Eq. (\ref{eq:rad_bary}) takes the form
\begin{align}
\nonumber\dot{M}_{b+r} &= (5.5\times10^{-34}\text{ g$\,$s$^{-1}$})\Biggl[0.25\rho_r'\biggl(\frac{1+z}{1+z_{mr}}\biggr)^4\\
&\phantom{==}+0.21\rho_b'\biggl(\frac{1+z}{1+z_{mr}}\biggr)^3\Biggr]M_{15}^2,\label{eq:rad_bary_mr}
\end{align}
\noindent so the accretion of the baryonic radiation fluid occurs slowly for near critical PBHs. This rate becomes comparable in magnitude to the Hawking evaporation rate when the hole is of characteristic mass
\begin{align}
\nonumber M_{ch,3b} &\simeq (2.1\times10^{22}\text{ g})\Biggl[0.25\rho_r'\biggl(\frac{1+z}{1+z_{mr}}\biggr)^4\\
&\phantom{==}+0.21\rho_b'\biggl(\frac{1+z}{1+z_{mr}}\biggr)^3\Biggr]^{-1/4}.\label{eq:M_ch3b}
\end{align}
\indent The accretion of dark matter onto a PBH will be horizon-limited and should be quite small if the spatially-averaged cosmological value for $\rho_{\chi,0}$ is assumed. The dark matter accretion rate is
\begin{align}
\dot{M}_\chi &= 4\pi R_S^2c\beta_\chi\rho_\chi,\label{eq:dmat_acc}
\end{align}
\noindent where $\beta_\chi$ is defined in Eq. (\ref{eq:v_rms}). The mass density of dark matter evolves according to $\rho_\chi = \rho_{\chi,0}(1+z)^3$ such that Eq. (\ref{eq:dmat_acc}) becomes
\begin{align}
\dot{M}_\chi &= \frac{16\pi G^2}{c^3}\frac{(\Theta_\chi^2+2\Theta_\chi)^{1/2}}{1+\Theta_\chi}\rho_{\chi,0}(1+z)^3M^2.\label{eq:dmat_acc2}
\end{align}
\noindent If there is an enhancement of the DM density term $\rho_\chi$ due to the formation of a DM halo there will be an appropriate enhancement of the DM accretion rate. Thus Eq. (\ref{eq:dmat_acc2}) represents a lower limit on the DM accretion rate. For a treatment of accretion from an enhanced DM halo see \cite{Mack07,Rico08}. Since the temperature $T_\chi$ of dark matter decoupled from the radiation temperature at $z_{fr}\sim2.1\times10^{13}$, the dimensionless quantity $\Theta_\chi$ in this redshift regime is quite small. The expansion of $\beta_\chi$ for $\Theta_\chi\ll 1$ is $\beta_\chi\simeq (2\Theta_\chi)^{1/2}$. Thus Eq. (\ref{eq:dmat_acc2}) becomes
\begin{align}
\nonumber\dot{M}_\chi &= \frac{16\pi G^2}{c^3}\biggl[\frac{3k_BT_0}{m_\chi c^2(1+z_{fr})}\biggr]^{1/2}\rho_{\chi,0}(1+z)^4M^2\\
&= (3.4\times10^{-58}\text{ g$\,$s$^{-1}$})(1+z)^4M_{15}^2,\label{eq:dmat_acc3}
\end{align}
\noindent which is about ten orders of magnitude smaller than the accretion rate due to the baryon-radiation coupled fluid. In this regime the accretion rate of dark matter onto a PBH becomes similar to the Hawking evaporation rate when
\begin{align}
M_{ch,3c} &= (2.3\times10^{28}\text{ g})(1+z)^{-1}.\label{eq:M_ch3c}
\end{align}
\noindent The constraints on the accretion rates further strengthens the argument that accretion onto a critical mass PBH is unimportant and most if not all of the lifetime of such a PBH is dominated by the Hawking evaporation.

\subsection{\label{subsec:DM_freeze}Post-DM freeze-out accretion}
\noindent In the post-DM freeze-out ($z_{mr}<z<z_{fr}$) era the universe is dominated by radiation. The dark matter, if it comprised of WIMPs, will be non-relativistic until redshifts higher than $z_{fr}$ \cite{Feng10} and will accrete at a horizon-limited rate. The accretion of baryonic matter and radiation is horizon-limited as before. It is convenient to apply $\rho_b\ll\rho_r$ and therefore ignore the baryonic matter terms and allow $c_s\sim c/\sqrt{3}$. Also in this redshift regime, the effective number of relativistic degrees of freedom $g_\star$ begins to increase at higher redshift so it is important to express the radiation term as in Eq. (\ref{eq:rho_rad}). The accretion rate is therefore
\begin{align}
\nonumber\dot{M}_{b+r} &= 4\pi R_S^2c_s(\rho_r+\rho_b)\\
\nonumber &\simeq \frac{8\pi^3 G^2k_B^4T_0^4}{15\sqrt{3}c^8\hbar^3}g_\star(1+z)^4M^2\\
&= (2.0\times10^7\text{ g$\,$s$^{-1}$})\biggl(\frac{g_\star}{86.25}\biggr)\biggl(\frac{1+z}{1+z_{fr}}\biggr)^4M_{15}^2.\label{eq:rad_bary_fr}
\end{align}
\noindent For a near critical mass PBH this is a large accretion rate compared to the magnitude of the Hawking rate. Thus the mass of a PBH in this redshift regime where these two rates balance is
\begin{align}
M_{ch,4} &= (1.5\times10^{12}\text{ g})\biggl(\frac{g_\star}{86.25}\biggr)^{-1/4}\biggl(\frac{1+z}{1+z_{fr}}\biggr)^{-1}.\label{eq:M_ch4}
\end{align}
\noindent The period of enhanced accretion in the early universe is quite short due to the strong redshift dependence, i.e. $\dot{M} \propto (1+z)^4$ so no significant accretion is expected for critical mass PBHs. This is consistent with the findings from previous studies on PBH accretion, i.e. \cite{Carr81}. At high redshift a critical mass PBH will not accrete significantly, but massive PBHs can grow by about an order of magnitude by $z_{mr}$.\\
\indent In this redshift regime the accretion of DM onto the PBH is small. It is increasingly important at higher redshift but is never larger than the radiation accretion rate in Eq. (\ref{eq:rad_bary_fr}). At the DM freeze-out redshift the DM particles are somewhat relativistic, i.e. $\Theta_\chi\sim 0.075$, such that the accretion rate is of the same form as Eq. (\ref{eq:dmat_acc3}) to a good approximation. In this regime Eq. (\ref{eq:M_ch3c}) also remains valid.

\subsection{\label{subsec:pre_DM_freeze}Pre-DM freeze-out accretion}
\noindent In this redshift regime the universe undergoes many changes as $g_\star$ increases and all particles become relativistic. At high enough redshifts all particles have the same temperature and follow $T=T_0(1+z)$. The accretion rate at these high redshifts is therefore the same as Eq. (\ref{eq:rad_bary_fr}). The PBH will not accrete radiation in the early universe if $T_{BH}>T_r$, which corresponds to $z<4.5\times10^{10}$ if $M = 10^{15}$ g. The radiation accretion at these high redshifts is highly dependent on the particle physics model. This study employs the Standard Model with all the latest particle masses from \cite{Oliv14}. The equivalent mass density in radiation changes dramatically in the early universe because of the change in $g_\star$ as shown in Table \ref{tab:particles}.\\
\indent Table \ref{tab:summary} summarizes the relevant properties of the universe with reference to the equations they are first noted. Table \ref{tab:M_dot} summarizes the relevant evaporation and accretion rates of PBHs in the relevant redshift regimes with reference to the equations or sections they are first noted. 
\begin{table}[t]
\begin{center}
\caption{\label{tab:summary} Properties of the universe across a large range in redshift.}
\begin{threeparttable}[hp]
\renewcommand{\arraystretch}{1.2}				
\begin{tabular}{lcccccc}
\toprule
$z$ & $\Omega_r$ & $\Omega_m$ & $\Omega_\Lambda$ & $T_r$ & $T_b$ & $c_s/c$\\
\hline
$z > z_{fr}$ & D\tnote{a} & N\tnote{b} & N & $\propto(1+z)$ & $=T_r$ & $\sim3^{-1/2}$\\
$z_{mr} < z < z_{fr}$ & D & N & N & $\propto(1+z)$ & $=T_r$ & $\sim3^{-1/2}$\\
$z_{rec} < z < z_{mr}$ & I\tnote{c} & D & N & $\propto(1+z)$ & $=T_r$ & Eq. (\ref{eq:sound_rad_bary})\\
$z_{th} < z < z_{rec}$ & N & D & N & $\propto(1+z)$ & $=T_r$ & Eq. (\ref{eq:bary_sound_post})\\
$z \lesssim z_{th}$ & N & D & N & $\propto(1+z)$ & Eq. (\ref{eq:baryon_temp})\tnote{e}& Eq. (\ref{eq:baryon_sound})\tnote{e}\\
$z = 0$ & N\tnote{d} & I\tnote{d} & D\tnote{d} & $=T_0$\tnote{d} & Eq. (\ref{eq:baryon_temp})\tnote{e} & Eq. (\ref{eq:baryon_sound})\tnote{e}\\
\bottomrule
\end{tabular}
\begin{tablenotes}
\item[a]{Dominant component of energy content.}
\item[b]{Negligible component of energy content.}
\item[c]{Important; non-negligible but non-dominant.}
\item[d]{$\Omega_{r,0}\sim9\times10^{-5}$, $\Omega_{m,0}\sim0.3089$, $\Omega_{\Lambda,0}\sim0.6911$, and $T_0=2.72548$ K; see \cite{Plan15} and \cite{Fixs09}.}
\item[e]{Does not account for reionization around $z\sim9$ or effects due to structure formation.}
\end{tablenotes}
\end{threeparttable}
\end{center}
\end{table}

\begin{table}[t]
\begin{center}
\caption{\label{tab:M_dot} PBH accretion and evaporation properties across a large range in redshift. This table summarizes the findings of \S\ref{sec:Evapo}. In each redshift regime, the accretion rates change due to the changes in $\rho_i$, $T_r$, $T_b$, and $c_s$ as in Table \ref{tab:summary}.}
\begin{threeparttable}
\renewcommand{\arraystretch}{1.2}				
\begin{tabular}{lcccc}
\toprule
$z$ & $\dot{M}_{evap}$\tnote{a} & $\dot{M}_r$ & $\dot{M}_b$ & $\dot{M}_\chi$\tnote{b}\\
\hline
$z > z_{fr}$ & Eq. (\ref{eq:Hawk}) & \multicolumn{2}{c}{Eq. (\ref{eq:rad_bary_fr})} & ---\\
$z_{mr} < z < z_{fr}$ & Eq. (\ref{eq:Hawk}) & \multicolumn{2}{c}{Eq. (\ref{eq:rad_bary_fr})} & Eq. (\ref{eq:dmat_acc3})\\
$z_{rec} < z < z_{mr}$ & Eq. (\ref{eq:Hawk}) & \multicolumn{2}{c}{Eq. (\ref{eq:rad_bary})} & Eq. (\ref{eq:dmat_acc3})\\
$z_{th} < z < z_{rec}$ & Eq. (\ref{eq:Hawk}) & Eq. (\ref{eq:rad_late}) & Eq. (\ref{eq:rec_Bond_Edd}) & Eq. (\ref{eq:dmat_acc3})\\
$z \lesssim z_{th}$ & Eq. (\ref{eq:Hawk}) & Eq. (\ref{eq:rad_late}) & Eq. (\ref{eq:Bond_min}) & Eq. (\ref{eq:dmat_acc3})\\
$z = 0$ & Eq. (\ref{eq:Hawk}) & Eq. (\ref{eq:rad_late}) & Eq. (\ref{eq:Bond_min}) & Eq. (\ref{eq:dmat_acc3})\\
\bottomrule
\end{tabular}
\begin{tablenotes}
\item[a]{Since the Hawking evaporation rate $\dot{M}_{evap}\propto M^{-2}$, it is only relevant if $M\lesssim M_{cr}$. High mass PBHs evaporate long after $z=0$; see Eq. (\ref{eq:evaptime})}
\item[b]{Does not account for DM halo formation in the late universe due to structure formation. Inside a DM halo the effective mass of the PBH will be enhanced by a potentially large factor $M\rightarrow f_{halo}M$ and thus $\dot{M}_\chi\rightarrow f_{halo}^2\dot{M}_\chi$.}
\end{tablenotes}
\end{threeparttable}
\end{center}
\end{table}

\section{\label{sec:Results} Results}
\begin{figure}[t]
\begin{center}
\includegraphics[width=\columnwidth]{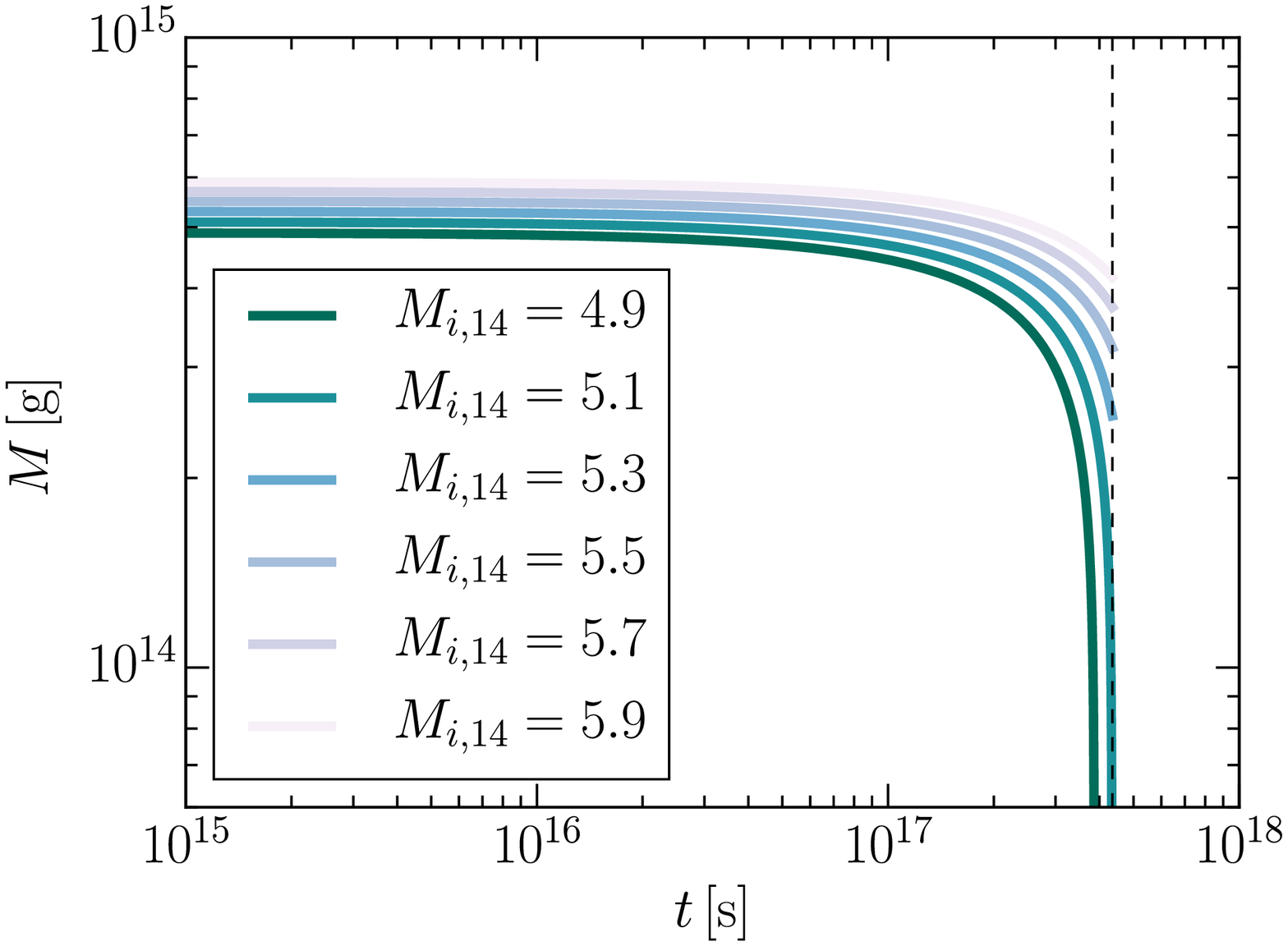}
\caption{\label{fig:waterfall} Waterfall plot of various PBHs forming at $z=10^{16}$ with masses near the critical evaporation mass, $M_{cr}\sim5.1\times10^{14}$ g. The PBHs near $M_{cr}$ suffer no significant accretion during their lifetime. The critical mass PBH evaporates at $t_{evap} = 13.8$ Gyr after the Big Bang (indicated by the dashed line). PBHs with $M < M_{cr}$ evaporate prior to the current era while those with $M > M_{cr}$ will evaporate in the future if they do not accrete significantly.}
\end{center}
\end{figure}

\begin{figure}[p]
\begin{center}
\includegraphics[width=\columnwidth]{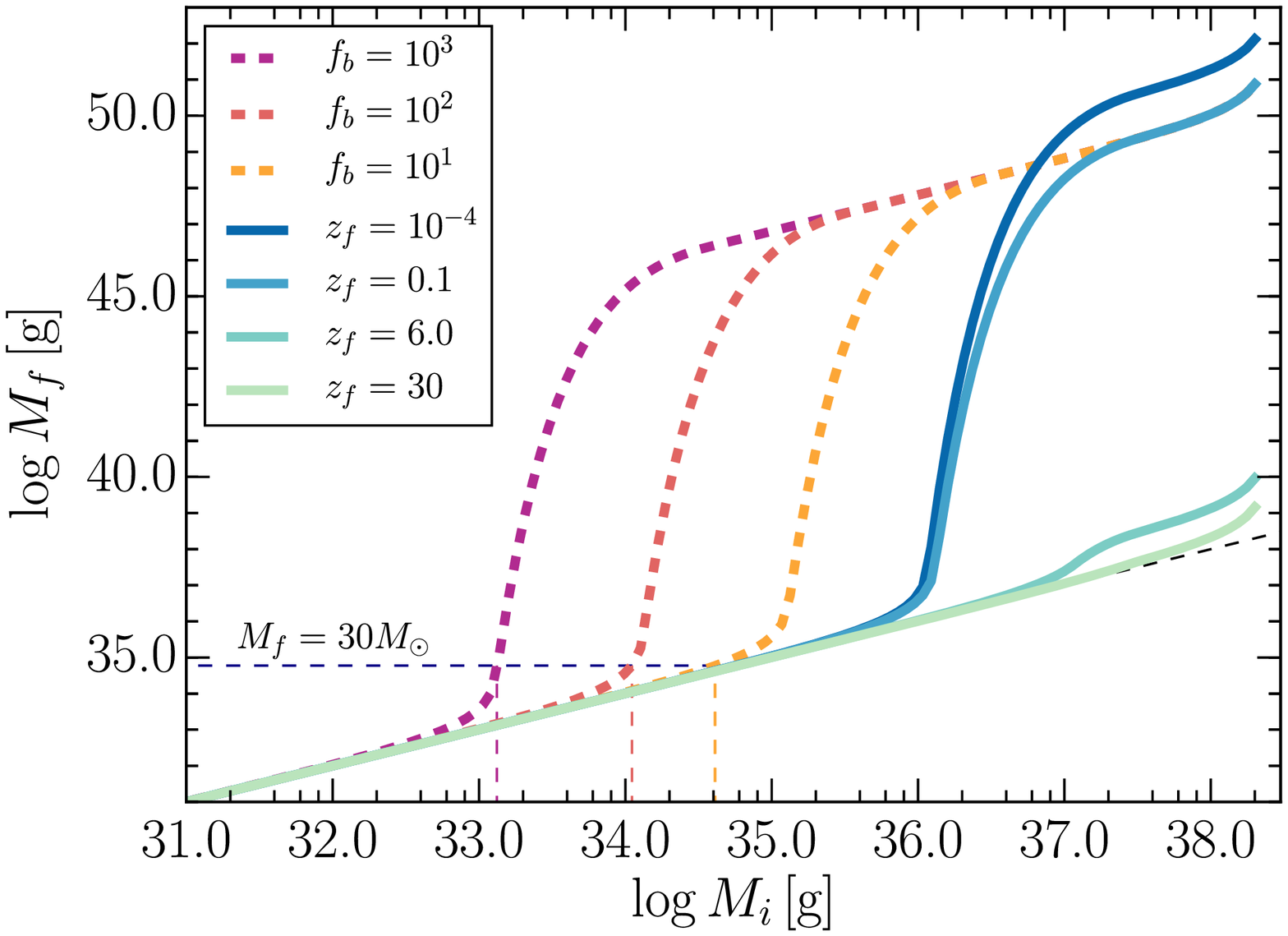}
\caption{\label{fig:lookup} Look-up plot of the final mass of PBHs forming at $1.0$ s after the Big Bang and ending at redshifts $30$, $6$, $0.1$, and $10^{-4}$. Also shown are three cases of PBHs forming at $1.0$ s after the Big Bang and ending at $z=0.1$ if they are located in a region where $f_b=10^1,10^2,$ and $10^3$. The plot shows the dramatic effects of late-universe accretion and density enhancement. It is known from SMBH observations that there are BHs with $M\sim2.5\times10^{43}$ g at $z\sim6.3$ \cite{Wu2015}. These holes are not easily explained with our `average' accretion histories; a PBH growing this large would have to be contained in an overdense region of the universe and supplied with gas for their entire histories. Laser Interferometer Gravitational-Wave Observatory (LIGO) observations of the gravitational wave events GW150914 \cite{Abbo16} and GW151226 \cite{Abb216} prove the existence of $\sim6\times10^{34}$ g and $\sim 3\times10^{34}$ g BHs at $z\sim0.1$. These observations are consistent with PBHs inside a regime of higher than average baryonic matter density that grow by a few orders of magnitude over their lifetime. The dotted vertical lines indicate the required initial masses that produce a PBH of $30M_\odot$ by $z=0.1$. Lower initial masses arise from higher density enhancements $f_b$.}
\end{center}
\end{figure}

\begin{figure}[p]
\begin{center}
\includegraphics[width=\columnwidth]{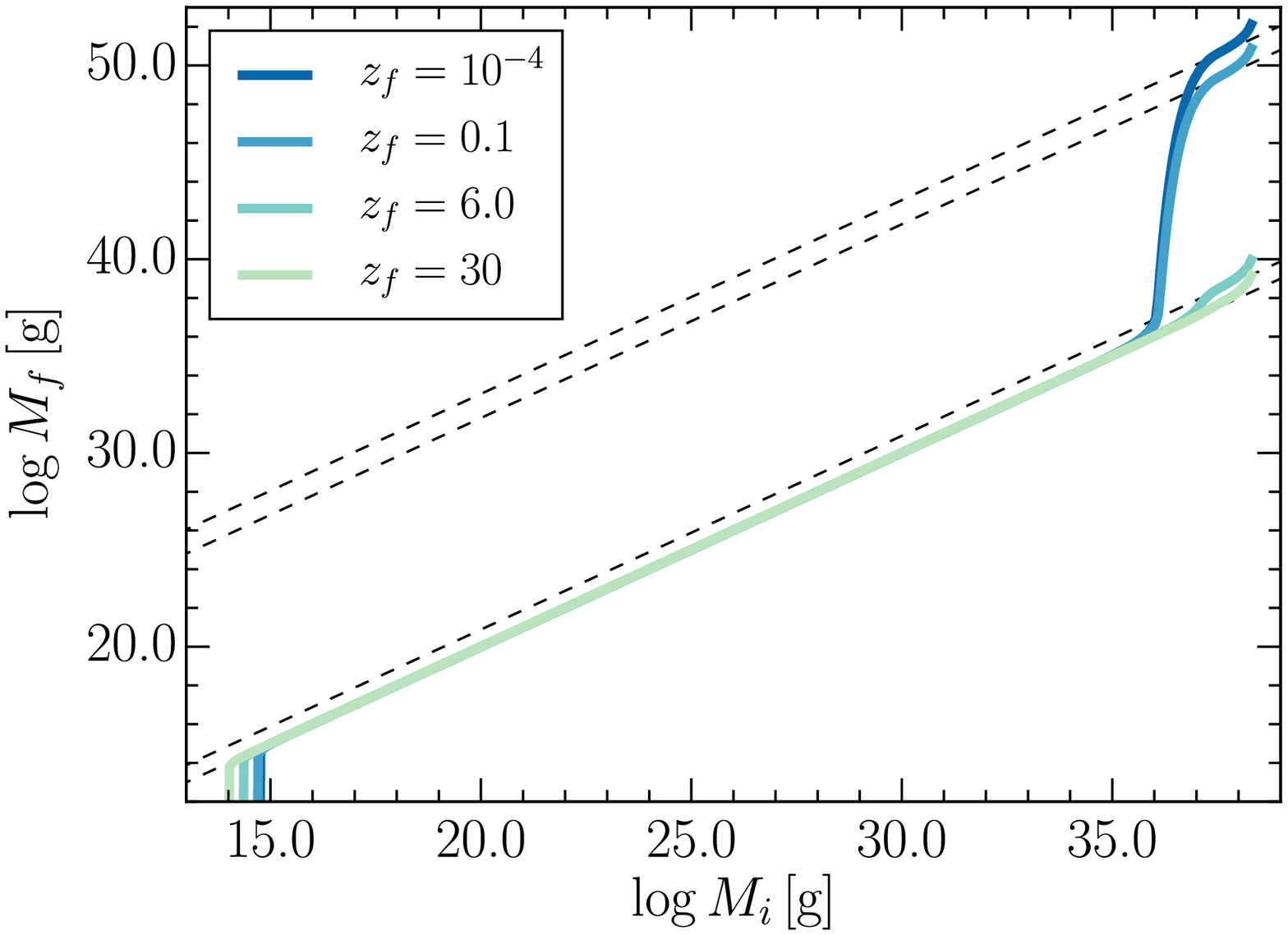}
\caption{\label{fig:final} Full look-up plot for all PBHs forming at $1.0$ s after the Big Bang and ending at redshifts $30$, $6$, $0.1$, and $10^{-4}$; same color scheme as Fig. \ref{fig:lookup}. The four dashed lines are $M_f=1.1\times10^{13}M_i$, $M_f=6.4\times10^{11}M_i$, $M_f=7.7M_i$, and $M_f=M_i$ (top to bottom). The $z_f=30$ case asymptotes to $M_f=1.01M_i$ for $M_i>10^{35}$ g. The low-mass regime agrees with Figure 2 of \cite{Carr16}, with the cut-off minimum mass increasing for lower final redshift (lower mass PBHs would have already evaporated). The increase in $M_f$ for $M_i\sim10^{38}$ g is due to the large accretion rate of Eq. (\ref{eq:rad_bary_fr}) at high redshift, which is large for only a short time due to the $(1+z)^4$ redshift dependence.}
\end{center}
\end{figure}

\noindent From the evaporation and accretion expressions in \S \ref{sec:Evapo} it is possible to construct a rough accretion or evaporation history for any PBH with mass $M_i$ forming at redshift $z_i$. The critical mass holes with $M_i = M_{cr} \sim 5.1\times10^{14}$ g will suffer no significant accretion in their entire lifetime if located in a suitably `average' volume of the universe. They will assume the evaporation timescale in Eq. (\ref{eq:evaptime}) and evaporate according to Fig. \ref{fig:waterfall}.
\\
\indent If the same PBHs of Fig. \ref{fig:waterfall} happened to form later, say at redshift $z_i = 10^8$, it would not affect their history due to the small timescales in the early universe. The accretion rate of the cosmological fluid onto larger PBHs at high redshift will be more important.\\
\indent The analysis in \S\ref{sec:Evapo} can be summarized in a look-up plot of $M_f$ against $M_i$. The regime important for $M_i\sim 10^{-4}M_\odot$ holes is shown in Fig. \ref{fig:lookup} and the entire mass regime is shown in Fig. \ref{fig:final}. Note the agreement of Fig. \ref{fig:final} in the near-critical mass regime to Figure 2 of \cite{Carr16}. The holes evaporating at higher redshift must have initial masses slightly lower than $M_{cr}$. Note that no significant accretion occurs across the intermediate mass regime between $M_{cr}$ and $\sim10^{36}$ g due to the low accretion rates for BHs of this mass. Isolated PBHs in this mass regime accreting the spatially-averaged cosmological fluid do not grow much. This does not account for enhancement of the accretion rates due to structure formation and thus represents a first approximation. If the accretion rate is enhanced via $\dot{M}_{b,B}\rightarrow f_b\dot{M}_{b,B}$ where $f_b = \rho_{enh}/\rho_b$ is an enhancement factor and $\rho_{enh}$ is the enhanced baryonic matter density, then a PBH of given initial mass can reach a higher mass for a given final redshift. This is reflected in the dotted lines of Fig. \ref{fig:lookup}, which show the final mass of a PBH growing from $1.0$ s after the big bang to $z=0.1$ given an enhancement factor $f_b=10^1,10^2,10^3$. Even a small enhancement of the baryonic matter density leads to a large increase in the possible final mass of the accreting PBH. Since the Bondi accretion rate is proportional to $M^2$, higher mass PBHs will accrete more than lower mass PBHs and this increase in the accretion rate is indicated by the increasing $M_f$ in Fig. \ref{fig:lookup} around $M\sim10^{36}$ g.\\
\indent In the first few seconds of the universe ($z\gtrsim10^{9})$, PBHs approaching the formation mass limit around $10^{38}$ g have a large accretion rate (see Eq. \ref{eq:rad_bary_fr}). This large accretion rate, though short-lived, can increase the mass of the PBH by about an order of magnitude by $z=10^9$. This effect is absent in lower mass PBHs and thus is visible in Fig. \ref{fig:final} as a small increase beginning above $M_i\sim10^{38}$ g.

\section{\label{sec:Discussion} Conclusions and Discussion}
\noindent A comprehensive view of the evolution of PBHs throughout cosmic history was presented. The accretion and evaporation histories of PBHs with masses in the approximate range $10^{14}\text{ g} < M < 2\times10^{38}\text{ g}$ were calculated. PBHs with lower masses will have evaporated prior to the current era and are not considered and PBHs with higher masses are not allowed due to the Hubble mass constraint of Eq. (\ref{eq:Hubble_mass}). The accreted fluids were assumed to have spatially averaged cosmological densities and the details of structure formation were not included. The important quantities for accretion are the mass densities of the various cosmological fluids, the sound speed in those fluids, and the details of their behavior at all relevant redshifts. All of these details were calculated precisely for the $\Lambda$CDM concordance cosmology.\\
\indent The important findings of this study are the following:
\begin{itemize}
\item A PBH with initial mass near $M_{cr}=5.1\times10^{14}$ g will not accrete radiation or matter in any significant quantity and will thus evaporate according to the timescale given in Eq. (\ref{eq:evaptime}). A PBH with initial mass less than $M_{cr}$ will evaporate prior to the current era.
\item A PBH with initial mass in the approximate range $10^{15}\text{ g} < M_i < 10^{35}\text{ g}$ neither evaporates nor accretes significantly over a Hubble time. Such a PBH would have to grow by other means, i.e. merging with other BHs or accreting while in an overdense region of the universe. Since the Hawking evaporation rate is so small for PBHs in this mass regime, the lower limit on the final (observed) mass of such PBHs is thus simply $M_f = M_i$.
\item A PBH with initial mass $M < 10^{38}$ g will not grow significantly in the early universe, i.e. within the first few minutes after the Big Bang. This finding is consistent with other PBH accretion studies, e.g. \cite{Carr81}. The small increase for BHs with $M_i\sim10^{38}$ g seen in Fig. \ref{fig:final} results from the large accretion rate for high-mass holes in Eq. (\ref{eq:rad_bary_fr}). It represents a growth of approximately one order of magnitude in the early universe, consistent with previous studies. There is negligible growth of critical mass PBHs in the radiation-dominated era.
\item A PBH with initial mass in the approximate range $10^{35}\text{ g} < M_i < 10^{37}\text{ g}$ can accrete significantly during its lifetime. In the redshift regime $z_{th}<z<z_{rec}$, a PBH with $M < (8.7\times10^{40}\text{ g})(1+z)^{-3/2}$ accretes at the Bondi rate and is Eddington-limited above that. In the redshift regime $z\le z_{th}$, a PBH with $M < 3.8\times10^{37}$ g accretes at the Bondi rate and is Eddington-limited above that. A PBH with such a mass that grows at the Bondi rate for its whole lifetime can thus grow by one or two orders of magnitude.
\item When a PBH grows enough to have its baryonic matter accrete at an Eddington-limited rate, the hole can increase in mass by many orders of magnitude if evolving into the late universe $z_f\sim 0$. Since the PBH will grow by accreting the spatially averaged cosmological gas, this growth represents how an `average' PBH accretes at the Eddington limit. The true accretion history of course will be complicated by feedback effects which were not modeled here. The curves in Fig. \ref{fig:final} thus represent an `average' growth. A true astrophysical hole of this mass may grow at either a higher or a lower rate.
\item A PBH with initial mass in the approximate range $4\times10^{37}\text{ g} < M_i < 10^{38}\text{ g}$ will accrete at an Eddington limited rate after $z_{rec}$ and the final mass of such a hole depends on its observed redshift. At $z_f=30$, the hole can only grow to $M_f=1.01M_i$. The hole can grow to $M_f=7.7M_i$ if $z_f=6$. The hole can grow to $M_f=6.4\times10^{11}M_i$ if $z_f=0.1$ and to $M_f=1.1\times10^{13}M_i$ if $z_f=10^{-4}$. See Fig. \ref{fig:final} for more details.
\end{itemize}
\noindent The PBH mass histories discussed in this study represent a first approximation of their cosmic behavior. Several astrophysical applications may be discussed in the context of the above results:
\begin{itemize}
\item It is impossible to explain the large BHs with $M\sim10^{10}M_\odot$ observed \cite{Wu2015} at $z>6$ via PBHs with Eddington-limited accretion of the `average' baryonic gas, even with $M_i\sim 10^5M_\odot$. These holes must be explained through multiple massive PBH mergers, mergers with BH seeds from the first generation of stars, or PBHs in overdense regions accreting at super-Eddington rates.
\item PBHs do not easily grow to $30M_\odot$ by $z\sim0.1$ through Bondi accretion of the `average' cosmological fluid. These PBHs cannot easily explain the binary BH mergers observed by LIGO as the gravitational wave events GW150914 \cite{Abbo16} and GW151226 \cite{Abb216} unless they experience an enhancement of the Bondi rate through various channels. One such channel is a baryonic matter density enhancement leading to $\dot{M}_{b,B}\rightarrow f_b\dot{M}_{b,B}$ as discussed in \S 4. Small enhancement factors allow a lower mass PBH to reach $30M_\odot$ compared to those PBHs accreting the average cosmological baryonic matter. Another possibility is the LIGO BHs were PBHs that formed with an initial mass $M_i=M_f$, where $M_f$ is their mass at the merger time. According to \cite{Sasa16}, the event rate for PBH mergers would be high enough to explain the GW events if the PBHs constitute a large enough fraction of the dark matter. However, PBHs in the appropriate mass range to explain these LIGO events are unlikely to be a large enough fraction of the DM as constrained from CMB measurements discussed in \cite{Chen16}, \cite{Horo16}, and \cite{AliH16}. Either LIGO has chanced upon two relatively rare PBH mergers or there is a common stellar evolution channel that produces BHs of these masses. Both explanations are interesting and more data are needed to distinguish these two possibilities.
\item Searches for PBH bursts \cite{Ukwa15} are ongoing. Although there are candidates for such events, no confirmed PBH burst event has been detected. The spectral properties of such bursts should be distinguishable from the `normal' GRBs. The non-detection of such an event has a few explanations. First, the fraction of PBHs that make up the dark matter must be quite low for PBHs of the relevant mass scale (see Fig. 9 of \cite{Carr10}). Thus it is plausible that not enough of these PBHs exist to subsequently evaporate and trigger gamma-ray detectors. Second, it might be possible for the critical mass holes to accrete enough to no longer evaporate in the current era. However, the accretion rate is too small and this would not explain the non-detection of PBH bursts. Even if the accretion rate onto small PBHs happened to be large enough, there would be smaller PBHs that would accrete enough to reach $M_{cr}$ anyway, filling the void of critical mass holes.
\end{itemize}

\section*{Acknowledgments}
\noindent This work was supported by the Nevada NASA Space Grant College and Fellowship Training Program Cooperative Agreement \#NNX10AN23H. We thank the anonymous referee for multiple suggestions that significantly improved the quality of the manuscript. J.R.R. thanks the UNLV High Energy Astrophysics Group for fruitful discussions. We extend our gratitude to Dr. Qing-Guo Huang of the Institute of Theoretical Physics, Chinese Academy of Sciences who alerted us to an error in an equation that is now fixed.

\appendix
\section{\label{Appendix_A} Maximum PBH formation mass}
\noindent As discussed in \S\ref{sec:Evapo} the maximum formation mass of a PBH will be the Hubble mass, i.e. the mass contained within the Hubble volume at a given time. The Hubble volume is
\begin{align}
V_H &= \frac{4}{3}\pi R_H^3,\label{eq:H_volume}
\end{align}
\noindent where $R_H = 2ct$ is the Hubble radius in the early radiation dominated universe. The critical density is the Hubble mass in a Hubble volume and thus
\begin{align}
\nonumber M_H(t) &= \rho_{cr}V_H\\
\nonumber&= \frac{3H^2}{8\pi G}\cdot\frac{4}{3}\pi(2ct)^3\\
&= \frac{4H^2c^3t^3}{G},
\end{align}
\noindent but in a radiation dominated universe $H=1/(2t)$ so that
\begin{align}
M_H &= \frac{c^3t}{G},
\end{align}
\noindent which recovers Eq. (\ref{eq:Hubble_mass}). See \cite{Carr10,Carr16} for more detailed discussions of the PBH mass function.

\section{\label{Appendix_B} Sound speed in the cosmological fluid}
\begin{figure}[t]
\begin{center}
\includegraphics[width=\columnwidth]{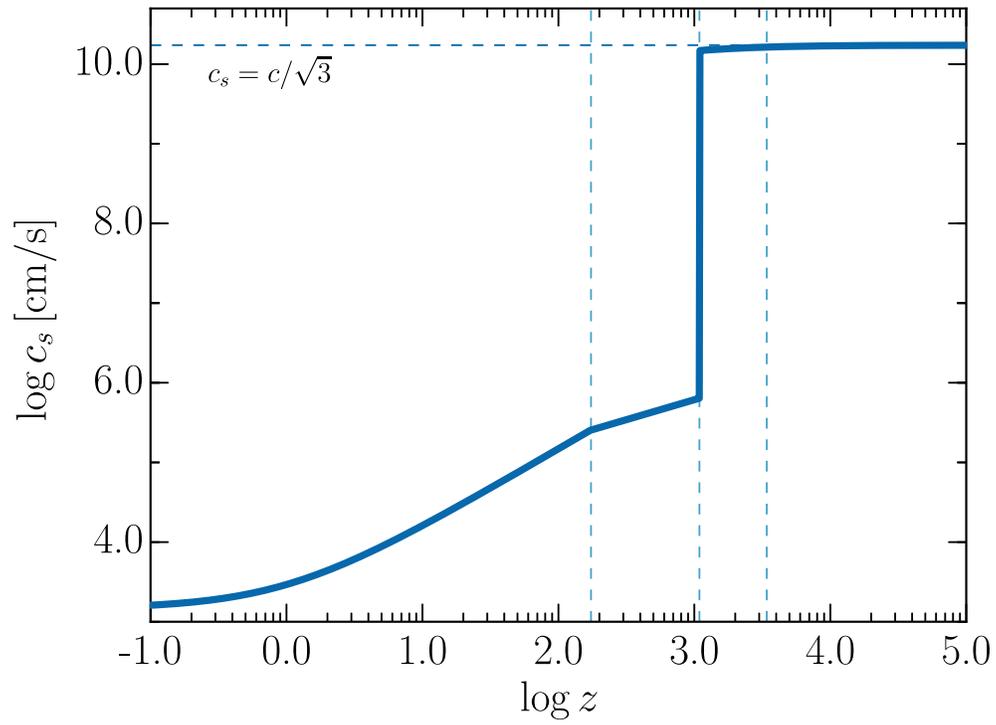}
\caption{\label{fig:cs} Plot of sound speed in the baryonic gas against redshift. The sound speed asymptotes to $c/\sqrt{3}\sim0.577c$ quickly after the recombination redshift. The three vertical dashed lines are (left to right) $z_{th}$, $z_{rec}$, and $z_{mr}$. The large jump at $z_{rec}$ is due to the decoupling of radiation and matter, which reduces the pressure.}
\end{center}
\end{figure}

\noindent In the late universe at redshifts lower than $z_{th}$, the temperature of the baryonic matter decouples from the CMB photon temperature. Thus the sound speed in the baryonic fluid is given by Eq. (\ref{eq:baryon_sound})
\begin{align}
c_{s,b} &= (1.5\times10^3\text{ cm$\,$s$^{-1}$})(1+z),
\end{align}
\noindent where the increase due to reionization around $z\sim9$ is not taken into account. In the redshift regime $z_{th}<z<z_{rec}$ the redshift dependence changes due to the temperature coupling between the baryonic matter and the CMB radiation. Thus the sound speed evolves as Eq. (\ref{eq:bary_sound_post})
\begin{align}
c_{s,b} &= (1.9\times10^4\text{ cm$\,$s$^{-1}$})(1+z)^{1/2}.
\end{align}
\noindent In the above equations it is assumed that the baryonic matter is composed entirely of hydrogen; corrections due to the helium and metal content of the baryonic matter need to be made for a more realistic calculation.\\
\indent In the early universe at redshifts higher than the recombination redshift $z_{rec}\sim1090$, the baryonic matter is coupled to the CMB radiation. The sound speed in such a coupled fluid can be found by calculating
\begin{align}
c_s^2 &= \biggl(\frac{\partial P}{\partial\rho}\biggr)_s,\label{A_eq:sound}
\end{align}
\noindent where the subscript $s$ on the right hand side indicates taking the derivative at constant entropy. The dominant pressure term is the radiation pressure and the density is a sum of radiation and baryonic terms $\rho = \rho_r + \rho_b$. The dark matter does not contribute to the pressure or density terms but has an early influence when it is relativistic at redshifts greater than $z_{fr}\sim2.1\times10^{13}$.\\
\indent Rewriting the partial derivatives of Eq. (\ref{A_eq:sound}) in terms of temperature gives
\begin{align}
c_s^2 &= \frac{(\partial P_r/\partial T)_s}{(\partial\rho_r/\partial T)_s+(\partial\rho_b/\partial T)_s}.\label{A_eq:sound2}
\end{align}
\noindent Recalling Eq. (\ref{eq:rho_rad}) and $P_r = \rho_r c^2/3$ the numerator of Eq. (\ref{A_eq:sound2}) is
\begin{align}
\biggl(\frac{\partial P_r}{\partial T}\biggr)_s &= \frac{4\pi^2}{90}g_\star(T)\frac{k_B^4T^3}{c^3\hbar^3} = \frac{4\rho_rc^2}{3T},\label{A_eq:rad_pressure}
\intertext{\noindent ignoring the small $\partial g_\star/\partial T$ terms. Similarly, the first term in the denominator of Eq. (\ref{A_eq:sound2}) is}
\biggl(\frac{\partial\rho_r}{\partial T}\biggr)_s &= \frac{4\pi^2}{30}g_\star(T)\frac{k_B^4T^3}{c^5\hbar^3} = \frac{4\rho_r}{T}.\label{A_eq:rad_density}
\end{align}
\noindent Recalling at high redshift the radiation and baryonic gas temperatures are coupled, i.e. $T_r = T_b = T$ and using $T = T_0(1+z)$, the second term in the denominator of Eq. (\ref{A_eq:sound2}) is
\begin{align}
\biggl(\frac{\partial\rho_b}{\partial T}\biggr)_s &= \frac{\partial}{\partial T}\biggl(\frac{\rho_{b,0}T^3}{T_0^3}\biggr) = \frac{3\rho_b}{T}\label{A_eq:baryon_density}.
\end{align}
\noindent Combining Eq. (\ref{A_eq:rad_pressure}--\ref{A_eq:baryon_density}) into Eq. (\ref{A_eq:sound2}) gives
\begin{align}
c_s^2 &= \frac{c^2}{3}\frac{4\rho_r}{4\rho_r+3\rho_b}.\label{A_eq:sound3}
\end{align}
\noindent It is clear that at redshifts higher than $z_{mr}\sim3400$ the sound speed calculated using Eq. (\ref{A_eq:sound3}) asymptotes to $c_s\sim c/\sqrt{3}\sim 0.577c$. The behavior of the sound speed across all relevant redshifts is plotted in Fig. \ref{fig:cs}.

\section*{References}
\bibliography{bib_PBH_long}

\end{document}